\documentclass[11pt]{article}

\usepackage{graphicx}
\begin{document}
\title{\bf Two Populations of Open Star Clusters in the Galaxy}

\author{{M.\,L.~Gozha, V.\,V.~Koval', V.\,A.~Marsakov} \\
{Southern Federal University, Rostov-on-Don, Russia}\\
{e-mail: gozhamarina@mail.ru, borkova@ip.rsu.ru, marsakov@ip.rsu.ru }}
\date{accepted \ 2012, Astronomy Letters, Vol. 38, No. 8, pp. 519-530}

\maketitle

\begin {abstract}

Based on our compiled catalogue of fundamental astrophysical parameters 
for 593 open clusters, we analyze the relations between the chemical 
composition, spatial positions, Galactic orbital elements, age, and 
other physical parameters of open star clusters. We show that the 
population of open clusters is heterogeneous and is divided into two 
groups differing by their mean parameters, properties, and origin. One 
group includes the Galactic clusters formed mainly from the 
interstellar matter of the thin disk with nearly solar metallicities 
($[Fe/H] > -0.2$) and having almost circular orbits a short distance
away from the Galactic plane, i.\,e., typical of the field stars of 
the Galactic thin disk. The second group includes the peculiar clusters 
formed through the interaction of extragalactic objects (such as 
high--velocity clouds, globular clusters, or dwarf galaxies) with the 
interstellar matter of the thin disk, which, as a result, derived 
abnormally low (for field thin-disk stars) metallicities and/or 
Galactic orbits typical of objects of the older Galactic subsystems. 
About 70\,\% of the clusters older than 1~Gyr have been found to be 
peculiar, suggesting a slower disruption of clusters with noncircular 
high orbits. Analysis of orbital elements has shown that the bulk of 
the clusters from both groups were formed within a Galactocentric 
radius of $\approx 10.5$ kpc and closer than $\approx 180$ pc from the 
Galactic plane, but owing to their high initial velocities, the 
peculiar clusters gradually took up the volumes occupied by the objects 
of the thick disk, the halo, and even the accreted halo of the Galaxy. 
Analysis of the relative abundances of magnesium (a representative of 
the $\alpha$-elements) in clusters that, according to their kinematical 
parameters, belong to different Galactic subsystems has shown that all 
clusters are composed of matter incorporating the interstellar matter 
of a single protogalactic cloud in different proportions, i.\,e., 
reprocessed in genetically related stars of the Galaxy. The [Mg/Fe] 
ratios for the clusters with thick-disk kinematics are, on average, 
overestimated, just as for the field stars of the so-called 
''metal--rich wing'' of the thick disk. For the clusters with halo 
kinematics, these ratios exhibit a very large spread, suggesting that 
they were formed mainly from matter that experienced a history of 
chemical evolution different from the Galactic one. We point out that a 
large fraction of the open clusters with thin-disk kinematics have also 
been formed from matter of an extragalactic nature within the last 
$\approx 30$ Myr.

\end{abstract}

{\em open star clusters, chemical composition, kinematics, Galaxy 
(Milky Way)}.

\maketitle

\section{Introduction}

In our previous paper (Gozha et\,al. 2012), we described the 
catalogue of fundamental parameters for open stars clusters that 
we compiled from the latest published data and discussed their 
errors and selection effects  . We showed the chemical and kinematical 
properties of clusters and field thin-disk stars to differ and 
provided evidence for the heterogeneity of the population of open 
clusters in the Galaxy. In this paper, we continue a comprehensive 
statistical analysis of the interrelations between the physical, 
chemical, and spatial--kinematical characteristics of open clusters 
and nearby field stars in order to reveal clusters of different 
natures and to estimate the relative number, characteristic parameters, 
and patterns in the identified populations of Galactic open star clusters.


\section {THE RELATION BETWEEN THE CHEMICAL COMPOSITION AND AGE OF OPEN CLUSTERS}

Marsakov et\,al. (1990) showed that the nearby field thin-disk 
stars and open clusters occupy slightly overlapping regions in 
the age--metallicity diagram and, in contrast to the field stars, 
a significant fraction of objects with a metallicity several times 
lower than the solar one are observed among clusters of any age. 
We will check the validity of this result based on currently 
available data. Since the stars that are currently located in the 
solar neighborhood were born at different Galactocentric distances, 
they reflect the chemical properties of the field thin-disk stars 
in a fairly wide range of distances. The mean or apogalactic radii 
of their orbits are most commonly assumed to be their most probable 
birthplaces. The mean orbital radii for the nearby field stars of our 
sample are 6 -- 10 kpc, while the apogalactic radii for some of them 
reach 13~kpc (see Fig.~2c below), i.\,e., they are comparable to the 
distances to the open clusters being investigated. Therefore, we have 
the opportunity to compare the chemical properties of nearby stars and 
open clusters far from the Sun. Figure 1a presents the distributions 
of the open clusters from our catalogue and the nearby (within 70~pc 
of the Sun) field thin-disk F-G dwarfs selected into this subsystem 
from the catalog by Holmberg et\,al.~(2009) based on the kinematical 
criteria from Koval' et\,al.~(2009) on the age-metallicity plane. 
The solid curve in the diagram indicates the cubic polynomial fit to 
the age-metallicity relation for the field stars (for a detailed 
justification and interpretation of the curve, see Marsakov et\,al.~2011).
The curve shows that the metallicity in the thin disk has increased 
appreciably within the last 4 -- 5 Gyr, while the mean metallicity at 
the initial evolutionary stages of this subsystem did not depend on the 
age, within the error limits. The polygonal line indicates the lower 
5\,\% envelope for the field stars, i.\,e., about 5\,\% of the stars 
lie below this line in the diagram at any age. This polygonal line is 
almost parallel to the average relation. As we see from Fig.~1a, a 
significant fraction of the clusters (open circles) occupy the upper 
left corner in the diagram, i.e., they are young and metal-rich. The 
ages of the three oldest clusters and the oldest field thin-disk stars 
are comparable in magnitude. We also see from the figure that a 
significant fraction of the clusters occupy a region almost completely 
free from field stars in the diagram at any age. As we see, clusters 
with metallicities in the range $[Fe/H] = (-0.4 \div -1.0)$ that is 
more typical of the thick-disk stars (see, e.g., Marsakov and Borkova 2005)
are being formed even at present. Note that mostly young stars are 
absent in our sample of  F--\,G dwarfs, because the errors in the 
isochrone ages of stars lying near the zero-age main sequence are often very
large and, therefore, they were not included in the sample. We extended 
the lower envelope toward younger ages (see the dotted line) based on
our analysis of the metallicity for more massive stars, Cepheids, 
whose ages are within the range 50 -- 300 Myr. According to the 
results from Andrievsky et al. (2002a), the mean metallicity of 
Cepheids in the solar neighborhood is $\langle[Fe/H]\rangle = +0.01$ 
and its dispersion is $\sigma_{[Fe/H]}=0.06$. Consequently, the 
metallicity of the 5\,\% lower point at zero age may be set equal 
to $-0.1$. We believe that it is the clusters lying below the polygonal 
line in the age-metallicity diagram rather than the clusters with the 
fixed limit $[Fe/H] < -0.2$, as was adopted by Vande Putte et\,al.~(2010),
that were formed from an interstellar medium with an appreciable 
admixture of matter with a different history of heavy--element 
enrichment. Out of the 264 clusters with known ages and metallicities, 
90 clusters satisfying this criterion were found in our catalogue.
In Fig.~1b, where the horizontal axis of the same diagram was 
transformed to a logarithmic scale, the dots inside open circles 
designate the clusters satisfying the criterion 
$(Z^2_{max}+4e^2)^{1/2} < 0.35$. We see that the overwhelming majority 
of such ''kinematically cold'' clusters lie above the lower envelope 
and are younger than 1~Gyr. However, there are clusters lying below 
the lower envelope among them as well, with almost all of these 
metal--poor clusters being younger than 30~Myr. We emphasize that 
allowance for the radial metallicity gradient did not distort the 
general appearance of the age--metallicity diagram, causing only some 
redistribution of field stars and clusters.

{\bf \em Relative Magnesium Abundances}. 

To analyze the chemical composition differences between the 
clusters, we used the best-studied $\alpha$--element, magnesium, 
because it has lines of different intensities and degrees of 
excitation in the visible spectral range in F--\,G stars. In the
age--[Mg/Fe] diagram (Fig. 1c), the open circles and the crosses 
indicate, respectively, our open clusters and the field thin--disk 
F--\,G stars from the catalogue by Borkova and Marsakov (2005). 
There are noticeable differences in the distributions of the field 
stars and clusters in the diagram, although both types of objects 
occupy approximately the same [Mg/Fe] range. As we see, the field 
stars show a clearly traceable sequence of the age--relative magnesium 
abundance relation. The corresponding correlation coefficients is 
$r = 0,37 \pm0,06$, and the probability of chance occurrence of the 
same correlation coefficient for the same number of points in the 
diagram is $P_N \ll 1$\,\%. Thus, the relative magnesium abundances in 
the thin--disk stars clearly increase with age. The correlation 
coefficient for the open clusters also turned out to be nonzero outside 
the error limits ($r = 0.27 \pm 0.14$, $P_N < 5$\,\%). However, in 
this case, as we see, the highest ratios $[Mg/Fe] > 0.25$ are observed 
in clusters younger than 4~Gyr, while these ratios in older clusters 
are considerably lower and they all lie in the band occupied by the 
field stars. Although there are only six clusters with such 
overestimated relative magnesium abundances in the diagram, they 
account for an appreciable fraction of all clusters with known 
[Mg/Fe] determinations. Therefore, it can be hypothesized that they 
could well be formed from matter that experienced a different history 
of chemical evolution than the interstellar medium of the thin disk. 
We also see from the figure that all ''kinematically cold'' clusters 
(except the very old Col 261) lie in the diagram in a region that, 
as it were, is an extension of the sequence of field F--\, stars toward 
younger ages (designated by the dots inside circles in the diagram).

{\bf THE RELATION BETWEEN THE CHEMICAL COMPOSITION AND POSITION IN THE GALAXY}.

When the evolution of elemental abundances in a stellar--gaseous 
system is studied, the total heavy--element abundance in the 
atmospheres of stars (metallicity) is most commonly used as the time 
axis. In this case, it is implied that in a closed system the heavy 
elements ejected into the interstellar medium by previous generations 
of stars lead to an inevitable increase in their abundances in younger 
stars, i.\,e., metallicity is a statistical age indicator. According 
to present views, the  $\alpha$-elements together with a small number 
of iron atoms are synthesized in massive ($M > 8 M_\odot$) asymptotic
giant branch stars that subsequently explode as Type II supernovae 
(SNe II) (Arnett 1978), while the bulk of the iron--peak elements 
are produced during Type Ia supernova (SN Ia) explosions 
(Thielemann et\,al.~1986). Starting from the paper by Tinsley (1979), 
the negative $[\alpha/Fe]$ trend as a function of metallicity observed 
in the Galaxy has been believed to be due to a difference in the 
evolution times of these stars. Indeed, the characteristic evolution 
time of SNe II is only $(10 \div 30)$~Myr, while numerous SN Ia 
explosions begin only after $1\div1.5$~Gyr (Matteucci and Greggio 1986; 
Yoshii et\,al.~1996). Therefore, having investigated the differences in 
relative $\alpha$--element abundances from cluster to cluster, we can 
attempt to ascertain how significant the genetic relationship between 
them is: the existence of a close correlation between metallicity and 
relative magnesium (a representative of the $\alpha$--elements) 
abundance in the system will provide evidence for a possible genetic 
relationship between the objects.

The [Fe/H]-[Mg/Fe] diagram for the open clusters (open circles) is 
shown in Fig.~1d. We see from this diagram that the spread in [Mg/Fe] 
for the clusters is much larger than that for the field thin-disk stars 
(crosses) at any metallicity. Therefore, the correlation for them also 
turned out to be insignificant ($r = -0.15 \pm 0.13$, $P_N =26$\,\%). 
However, the slope of the regression line for the clusters (solid line) 
was found to be the same, within the error limits, as that for the 
field stars (dashed line). As we see, the clusters of the kinematically 
cold group with known magnesium abundances (dots inside circles) lie 
mostly in the band occupied by the field stars (they all are equally 
rich in metals and, therefore, form no sequence by themselves). Note 
that the errors of the relative magnesium abundances in open clusters 
usually declared by the authors ($\epsilon [Mg/Fe]_{\textrm{ск}} = \pm 0.08$)
are not much larger than those for single nearby field stars 
($\epsilon[Mg/Fe]_{\textrm{зв}} = \pm 0.05$), because the corresponding 
values were obtained for each cluster by averaging over several stars 
(for more details, see Gozha et\,al.~2012). Therefore, the large spread 
in [Mg/Fe] for the clusters cannot be attributed to their large errors. 
Thus, despite the large spread in relative magnesium abundances at a 
given metallicity, the existence of a weak correlation between [Mg/Fe] 
and [Fe/H] for the open clusters suggests that some genetic 
relationship most likely exists between the clusters.

{\bf \em The Radial and Vertical Metallicity Gradients} 

Let us now consider how the mean metallicity of the clusters changes 
with distance from the Galactic center, i.\,e., let us trace the radial 
metallicity gradient. Figure~2a presents the Galactocentric distance 
($R_G$)--metallicity diagram for all clusters of our catalogue. For a 
more detailed examination of the diagram, we restricted it to the range 
$5 \pm 18$ kpc, with only one cluster being to the left and three 
clusters reaching $\approx22$~kpc being to the right. However, these 
points are taken into account in all calculations. The dashed line 
constructed from all of the observed open clusters is the regression 
line and gives a gradient $d[Fe/H]/dR_G = (-0.043 \pm 0.006)$~kpc$^{-1}$
(r = -0.40 ± 0.05, PN << 1
other authors (see, e.g., Chen et\,al.~2003,~2008; Wu et\,al. 2009; 
Andreuzzi et \,al. 2011). The large circles designate the mean 
metallicities in narrow $R_G$ ranges, and the bars designate their rms 
deviations. The polygonal curve connecting these points shows the 
pattern of change in metallicity with distance from the Galactic 
center in more detail than does the regression line. We see that 
the mean metallicity decreases abruptly by ?[Fe/H] = -0.25 when 
passing through $R_G \approx 9.5$~kpc, while the metallicity before 
and after this distance may be considered unchanged, within the error 
limits. The dots inside circles highlight the clusters obeying the 
condition ($Z^2_{max}+4e^2)^{1/2} < 0.35$ , i.\,e., the ''kinematically 
cold'' clusters. We see that the step-like pattern of the 
''Galactocentric distance--metallicity'' relation stems from the fact 
that most of such clusters turn out to be metal--rich ($[Fe/H] > -0.2$) 
and to lie within 9 kpc of the Galactic center (see the concentration 
in the diagram), while the distant clusters turn out to be mostly less 
metal--rich. (However, it should be noted that since all ''kinematically
cold'' clusters are near the Galactic plane, they may not be identified 
beyond this radius due to strong interstellar extinction.) For 
comparison, the crosses in the diagram indicate the distant visible 
young thin-disk stars, Cepheids, from Andrievsky et\,al.~(2002a - 
2002c, 2004) and Luck et\,al.~(2003). We see from the diagram that the 
field Cepheids form a narrow sequence and exhibit a slightly larger 
radial gradient than do the clusters: 
($d[Fe/H]/dR_G)_{зв} = (-0.056 \pm 0.003$ kpc$^{-1}$. However, the 
radial gradient for the "kinematically cold" clusters lying, just as 
the Cepheids, in the same narrow layer near the Galactic plane turned 
out to be larger than that for the Cepheids outside the error limits: 
$d[Fe/H]/dR_G = (-0.10 \pm 0.02$ kpc$^{-1}$ ($r = -0.37 \pm 0.08$, 
$P_N \ll 1$\,\% ). Note that almost all of the open clusters with ages 
of less than 300 Myr, i.\,e., as young as the Cepheids, lie within 
$\approx11$~kpc of the Galactic center. They show almost the same 
gradient as do the ''kinematically cold'' clusters: 
$d[Fe/H]/dR_G = (-0,09 \pm 0.02)$~kpc$^{-1}$ ($r = -0.37 \pm 0.08$,
$P_N \ll 1$\,\%). The older clusters exhibit a smaller gradient than 
do the young ones: $d[Fe/H]/dR_G = (-0,05 \pm 0.01)$~kpc$^{-1}$ 
($r = -0.49 \pm 0.07$, $P_N \ll 1$\,\%). Note that this result is in 
conflict with the steeper radial gradient usually obtained from old 
open clusters than that for young ones (see, e.g., 
Magrini et\,al.~(2009) and references therein). The steep gradient 
results from the restriction of the range of Galactocentric distances 
by $<12$~kpc, i.\,e., if we ignore the fact that the more distant 
clusters exhibit a higher metallicity than do the clusters lying in 
the range $(9 \div 12)$~kpc (see Fig. 2a).

Consider the situation with the vertical metallicity gradient for the 
same objects. Figure 2b presents their distance from the Galactic plane 
(taken in absolute value)-metallicity diagrams. The regression line 
points to the existence of a steep negative vertical metallicity 
gradient for the clusters: $d[Fe/H]/dz = (-0.17 \pm 0.05)$~kpc$^{-1}$ 
($r = -0.23 \pm 0.06$, $P_N \ll 1$\,\%). Approximately the same 
(within the error limits) gradients are obtained for both field 
Cepheids and ''kinematically cold'' clusters (the corresponding 
regression lines are not shown). However, the average line points to 
the existence of an abrupt decrease in metallicity by $\Delta[Fe/H]= -0.10$
when passing through $|z|\approx 180$~pc (although the jump here is 
considerably smaller in magnitude than that for the radial gradient, 
but, as we see from the figure, it also manifests itself outside the 
limits of the errors in the mean points). Just as in the preceding 
diagram, the concentration of points formed almost completely by the 
''kinematically cold'' clusters is clearly distinguished here 
($[Fe/H] > -0.2$; $|z| < 100$~ pc).

Thus, the $R_G - [Fe/H]$ and $|z| - [Fe/H]$ diagrams show the same 
morphological structure that consists in the presence of concentrations 
of metal--rich clusters and abrupt decreases in metallicity when passing
from these concentrations to greater distances. This most likely 
suggests that both negative metallicity gradients are due to the 
existence of two types of populations with differing metallicities and 
spatial distributions among the open clusters. Our checking showed that 
the results and conclusions also remain valid if not the present 
locations of the clusters but the maximum distances of their orbital 
points from the Galactic center and plane are used as the cluster 
distances from the Galactic center and plane. Recall that 
Yong et\,al.~(2005) and Magrini et\,al.~(2009) point to a decrease in 
the slope of the radial metallicity gradient when passing through 
12~kpc. However, we believe that it is still more appropriate to talk 
about the metallicity jump at a shorter distance equal to 
$\approx 9.5$~kpc. This follows from the fact that, first, a similar 
step--like pattern is also observed for the vertical metallicity 
gradient and, second, both jumps are naturally explained by the same 
factor -- the existence of two approximately equal (in number) 
populations of open clusters (see below).

{\bf \em The Radial and Vertical Gradients in Relative Magnesium Abundances} 
Figure 2c presents the $R_G - [Mg/Fe]$ diagram for the open clusters 
(open circles). We see that because of the large spread in [Mg/Fe] 
among the clusters, the correlation between their Galactocentric 
positions and the magnesium abundance turns out to be insignificant 
($r = 0.21 \pm 0.13$, а $P_N \ll 1$\,\%), although the radial gradient 
in relative magnetic abundance slightly differs from zero outside the 
error limits, $d[Mg/Fe]/dR_G = (0.007 \pm 0.005$)~kpc$^{-1}$.(If, 
however, the maximum orbital radii are used for the clusters, then the 
gradient disappears altogether.) Because of the small number of 
magnesium abundance determinations in the clusters, a statistically 
significant existence of the jump in the relation cannot be traced, but 
the average line shows a clearly traceable break in the relation near 
$R_G \approx 9.5$~kpc. Note that all ''kinematically cold'' clusters 
exhibit low relative magnesium abundances, $[Mg/Fe] < 0.15$. No radial 
magnesium gradient is traceable ($P_N = 75$\,\%) from our sample of 
field thin--disk stars, which are indicated in the same panel by the 
crosses for comparison. Since all stars in this catalogue are nearby 
ones, we had to use the apogalactic radii of their orbits by assuming 
that they were formed precisely at these distances.

Figure 2d presents the $|z| - [Mg/Fe]$ diagram for the open clusters. 
The correlation here turned out to be also insignificant 
($r = 0.23 \pm 0.13$ andи $P_N = 9$\,\%), although a small slope is 
also traceable: $d[Mg/Fe]/dz =(0.05 \pm 0.03)$~kpc$^{-1}$. Just as in 
Fig. 2b, the average line here exhibits a jump, but its amplitude does 
not go outside the errors of the means and it is observed at a larger 
$|z|$. For the nearby field thin--disk stars, the vertical gradient in 
relative magnesium abundance turns out to be slightly larger: 
($d[Mg/Fe]/dZ_{max})_{\textrm{st}} = 0.11 \pm 0.03)$~kpc$^{-1}$ 
($r = 0.24 \pm 0.06$ and $P_N \ll 1$\,\%. The ''kinematically cold'' 
clusters that we identified show no correlation, because, on the one 
hand, they cannot be at large distances from the Galactic plane by 
definition and, on the other hand, a significant spread in relative 
magnesium abundances is observed among them.

{\bf \em THE STRATIFICATION OF OPEN CLUSTERS INTO POPULATIONS}. 

The Galactic field stars are known to show correlations between their 
spatial positions, orbital elements, chemical compositions, and ages. 
These correlations arose from significant differences between all of 
the listed parameters for four Galactic subsystems: the thin disk, the 
thick disk, the halo, and the accreted halo, i.\,e., corona (see, e.g., 
Marsakov and Borkova 2005, 2006a, 2006b). The first three subsystems 
are genetically related, i.\,e., their objects were formed from the 
interstellar matter of a single protogalactic cloud. The differences 
in parameters between these subsystems are attributable to the 
chemical and dynamical evolution of the interstellar matter of this 
collapsing protocloud. In contrast, the stars of the fourth subsystem 
were born in dwarf satellite galaxies that were subsequently disrupted 
under the tidal forces of the Galaxy. These stars were formed from 
matter that experienced chemical evolution different from the Galactic 
one (Marsakov and Borkova 2006b).
In contrast to the field stars, all open clusters were formed from the 
interstellar matter of the thin Galactic disk and, thus, all of them 
are initially composed (at least partially) from the matter of a single 
protogalactic cloud. However, the properties of the open clusters 
described above unequivocally point to the heterogeneity of this 
population in the Galaxy. On the one hand, two groups can be identified 
among them according to their spatial-kinematical parameters. The first 
group includes the clusters with nearly circular orbits a short 
distance away from the Galactic plane (just as for the young population 
of the thin disk), while the second group includes the clusters with 
highly eccentric and high orbits typical of the thick--disk and even 
halo objects. On the other hand, the clusters can also be divided into 
two groups by the total heavy-element abundance in their stars. The 
bimodality of their metallicity distribution (see Gozha et\,al.~(2012) 
and references therein) and their positions in the age--metallicity 
diagram above and below the lower envelope for the field stars 
(see Figs. 1a and 1b) are the main prerequisites to this division. 
Although the groups in metallicity and Galactic orbital elements 
mutually overlap, there is no close correspondence between them. 
Therefore, we will separate the clusters with a problematic origin 
independently by these two parameters. The clusters with low circular 
orbits (i.e., satisfying the criterion ($Z^2_{max}+4e^2)^{1/2} < 0.35$) 
and simultaneously lying above the lower envelope for the field stars 
in the age-metallicity diagram will be called Galactic or ''thin--disk''
ones. Conversely, the clusters with atypically low metallicities for 
the thin--disk stars and the clusters with eccentric high orbits will 
be called peculiar ones. To this group we added 31 clusters for which 
no orbital elements were determined but they are currently at a 
distance $|z| > 0.35$~kpc from the Galactic plane, definitely 
satisfying the kinematical criterion used above. By our definition, the 
clusters for which either the orbital elements or the metallicities 
were not found also fall into the peculiar group.

Out of the 264 open clusters with known metallicities, 90 turned out to 
have [Fe/H] atypically low for the thin--disk stars. At the same time, 
30 such metal-poor clusters have kinematics like that for the thin--disk
stars, while 17 clusters have orbits typical of the Galactic thick disk 
and halo. There are no orbital elements for 43 metal-poor clusters, 
but 13 of them have $|z| > 0.35$~kpc. Twenty seven clusters with high 
metallicities have eccentric high orbits and 18 more clusters lie far 
from the Galactic plane. The peculiar group contains a total of 182 
clusters. The spread in orbital parameters among the peculiar clusters 
turned out to be so large that we separated the clusters with 
characteristic ($(Z^2_{max}+4e^2)^{1/2} > 2.3$, i.\,e., exceeding the 
mean value for the group of peculiar clusters almost by $3\sigma$, into an
individual group. The bulk of the peculiar clusters constitute the 
''thick-disk'' group, while six clusters were separated into the 
individual ''halo'' group -- Berkeley 20, Berkeley 21, Berkeley 29, 
Berkeley 31, Berkeley 33, and Berkeley 99. The cluster Berkeley 29 
turned out to have retrograde rotation around the Galactic center, and 
this unequivocally points to an extragalactic nature of the object. 
Clearly, the separation of clusters into the Galactic subsystems is 
arbitrary in the sense that these three groups only occupy the regions 
in the Galactic space that correspond to its subsystems of the same 
name. In fact, however, all of them, of course, were formed at 
different times in the thin disk mainly from its interstellar matter, 
which derived an admixture of matter with a different chemical 
composition and an additional momentum. The latter, on the one hand, 
led to star formation and, on the other hand, imparted an initial 
velocity to the formed cluster. Depending on the nature of the source 
of this momentum, the orbits and chemical compositions of the clusters 
will differ. The hypothesis that these clusters were formed at the 
maximum distances of their orbital points from the Galactic plane seems 
less probable, because the density of interstellar matter there is an 
order of magnitude lower than that within several tens of parsecs and 
star formation is unlikely to be possible at such a density. The 
membership of clusters in the ''thin--disk'', ''thick--disk'', and 
''halo'' populations is denoted in our catalogue by 1, 2, and 3, 
respectively (see Gozha et\,al.~2012), while the characteristic 
parameters for the identified populations of open clusters are listed 
in the table, where the number of clusters used to determine the 
parameters is given in parentheses.

\begin{table}
\centering
\caption{%
  Характерные параметры трех населений рассеянных скоплений}
\begin{tabular}{|l|c|c|c|c|c|}
\hline
\multicolumn{1}{|c|}{\bf Parameters} & 
\multicolumn{1}{c|}{\bf Galactic} &
\multicolumn{2}{c|}{\bf Peculiar clusters} &
\multicolumn{2}{c|}{\bf Neaarby clusters($d_\odot <1$~kpc)}\\
\cline{2-6}
& {\bf ''thin disk''} & {\bf''thick disk''} & {\bf ''halo''} &
{\bf ''thin disk''}  & {\bf ''thick disk''}\\
\hline

$\langle[Fe/H]\rangle             $&$0.03\pm0.01 (96)$&$-0.33\pm0.02(129)$&$-0.50\pm0.09(6)$&$0.03\pm0.02(55)$&$-0.36\pm0.10(7)$\\
$\sigma [Fe/H]                    $&$0.12\pm0.01     $&$0.26\pm0.02     $&$0.22\pm0.06$&$0.12\pm0.01$&$0.27\pm0.07$\\
$\langle[Mg/Fe]\rangle            $&$0.02\pm0.03(15) $&$0.11\pm0.02(31) $&$0.10\pm0.09(5)$&$0.020.04(9)$&$0.08(1)$\\
$\sigma [Mg/Fe]                   $&$0.11\pm0.02     $&$0.11\pm0.01     $&$0.20\pm0.06$&$0.12\pm0.03$&$-$\\
$t,\textrm{млрд.\,лет}            $&$0.40\pm0.10(95) $&$0.91\pm0.11(173)$&$2.55\pm0.78(6)$&$0.33\pm0.05(54)$&$0.72\pm0.38(11)$\\
$\sigma(t) \textrm{млрд.\,лет}    $&$0.95\pm0.07     $&$1.40\pm0.08     $&$1.90\pm0.55$&$0.38\pm0.0$&$1.25\pm0.27$\\
$\langle lg M/M_\odot\rangle      $&$2.58\pm0.07(82) $&$2.99\pm0.09(83) $&$-$&$2.46\pm0.09(54)$&$2.50\pm0.23(10)$\\
$\sigma [lg M/M_\odot]            $&$0.63\pm0.05     $&$0.85\pm0.07     $&$-$&$0.64\pm0.06$&$0.72\pm0.16$\\
$\langle ellipticity\rangle       $&$0.26\pm0.02(82) $&$0.35\pm0.02(83) $&$-$&$0.22\pm0.02(54)$&$0.24\pm0.05(10)$\\
$\sigma (ellipticity)             $&$0.18\pm0.01     $&$0.19\pm0.01     $&$-$&$0.13\pm0.01$&$0.15\pm0.03$\\
$\langle lg(r_{cl}/r_{co})\rangle $&$0.48\pm0.02(82) $&$0.38\pm0.01(83) $&$-$&$0.52\pm0.02(54)$&$0.46\pm0.05(10)$\\
$\sigma {lg(r_{cl}/r_{co})}       $&$0.15\pm0.01     $&$0.13\pm0.01     $&$-$&$0.14\pm0.01$&$0.15\pm0.03$\\
$\langle e \rangle                $&$0.07\pm0.01(96) $&$0.16\pm0.01(115)$&$0.46\pm0.09(6)$&$0.06\pm0.01(55)$&$0.14\pm0.03(10)$\\
$\sigma (e)                       $&$0.032\pm0.002   $&$0.11\pm0.01     $&$0.22\pm0.06$&$0.03\pm0.01$&$0.10\pm0.02$\\
$\langle Z_{max}\rangle           $&$0.11\pm0.01(96) $&$0.39\pm0.04(115)$&$7.71\pm2.72(6)$&$0.10\pm0.01(55)$&$0.15\pm0.04(10)$\\
$\sigma  (Z_{max})                $&$0.07\pm0.01     $&$0.42\pm0.03     $&$6.65\pm1.92$&$0.06\pm0.01$&$0.13\pm0.03$\\
$\langle R_{GC} \rangle           $&$7.86\pm0.08(96) $&$9.97\pm0.18(176)$&$15.14\pm1.65(6)$&$7.98\pm0.06(55)$&$8.01\pm0.17(12)$\\
$\sigma (R_{GC})                  $&$0.74\pm0.05     $&$2.45\pm0.13     $&$4.03\pm1.16$&$0.45\pm0.04$&$0.60\pm0.12$\\
$ Z_{o}                           $&$0.07\pm0.01(99) $&$0.27\pm0.03(173)$&$-$&$-$&$-$\\

\hline
\end{tabular}
\end{table}

In the table, we are primarily interested not in the absolute values 
of the parameters but only in the comparative characteristics of the 
populations of open clusters, because some of these quantities can be 
distorted by observational selection effects. To have an idea of how 
strongly they affect the mean parameters, we give the corresponding 
quantities for the ''thin''- and ''thick-disk'' clusters lying within 
1~kpc of the Sun in the two extreme columns, because the selection is 
minimal within these limits (Piskunov et\,al.~2006). Taking into 
account the selection for distant clusters and the very low reliability 
of the means for nearby peculiar clusters (because of their small number),
below we give only a qualitative description of the differences. Note 
that, according to the proposed stratification scheme, the number of 
peculiar open clusters in our catalogue turned out to be more than a 
factor of 1.5 larger than that of Galactic ones. However, such a high 
percentage of peculiar clusters in our sample was obtained due to the 
selection in favor of the clusters located at high Galactic latitudes, 
because the distances and metallicities are easier to determine for 
them. Indeed, among more than 2000 optically detected (Dias et\,al.~2002)
open clusters, the overwhelming fraction of them lies near the Galactic 
plane. As a result, the actual fraction of the peculiar clusters is 
unlikely to exceed a third of their total number.
The difference in metallicity turned out to be the expected one: the 
metallicity is nearly solar for the Galactic clusters, while it is 
approximately a factor of 2.5 lower for the ''thick--disk'' clusters. 
At the same time, the metallicity dispersion for the second group is 
much greater. At such a large difference in metallicity between the 
groups, we obtained the explainable difference in relative magnesium 
abundances toward higher values for the ''thick--disk'' clusters. At an 
even lower metallicity for the ''halo'' clusters, the mean relative 
magnesium abundance in this group did not increase but even slightly 
decreased compared to the thick disk (see the distribution of clusters 
of different populations in Fig.~3a). The mean age and age dispersion 
for the ''thick--disk'' clusters were found to be greater than those 
for the ''thin--disk'' clusters outside the error limits. Note that 
70\,\% of the clusters at $t > 1$~Gyr are peculiar. An even greater age 
(and an even larger age dispersion) was obtained for the ''halo'' 
clusters. Here, it should be remembered that the bulk of the old open 
clusters in each group have already been disrupted and the difference 
in ages points only to the degree of survivability of clusters with 
different orbits.
The mean mass and its dispersion for the ''thick--disk'' clusters 
turned out to be larger outside the error limits (however, the 
differences for nearby clusters are smaller, but these are most likely 
biased estimates, because there are data on the physical parameters 
only for ten peculiar clusters). In the RG-log(M/M ) diagram in Fig.~3b,
we can not only examine the mass distributions of ''thin''- and 
''thick--disk'' clusters but also trace the increase in the lower mass 
limit as one recedes from the solar Galactocentric distance in both 
directions. Such a behavior implies that the higher, on average, mean 
mass for the peculiar clusters was obtained due to observational 
selection - distant low-mass clusters are harder to distinguish against 
the background of field stars and the distances are more difficult to 
determine for them. Unfortunately, there are no physical parameters for 
the ''halo'' clusters.
The mean ellipticities of the clusters in the disk populations at 
identical dispersions show a difference outside the error limits - the 
''thick-disk'' clusters turn out to be more deformed. The distributions 
of ''thin''- and ''thick-disk'' clusters on the central 
concentration--ellipticity plane can be seen in Fig.~3c. The central 
concentrations turned out to be larger than those for the ''thin-disk'' 
clusters outside the error limits - the corresponding distribution 
exhibits a clear asymmetry toward high values $lg(r_{cl}/r_{co})$, i.e.,
dynamically, these clusters turn out to have evolved farther than the 
peculiar ones, although their ages are, on average, younger.
The mean eccentricities of the orbits and their maximum distances from 
the Galactic plane as well as the corresponding dispersions are 
greater, outside the error limits, for the ''thick--disk'' clusters. The
mean values of these parameters for the ''halo'' clusters are greater 
by several more times. This is undoubtedly a consequence of the 
selection into groups. However, in any case, the atypically large 
(for the thin Galactic disk) orbital elements for the peculiar clusters 
suggest a high energy of the factors that perturbed their motion. Since 
the clusters spend much of the time near the apogalactic radii of their 
orbits, it is not surprising that the mean Galactocentric distances for 
clusters with eccentric orbits turned out to be greater as well. The 
maximum in the $R_G$ distribution for the Galactic clusters is equal, 
within the error limits, to the solar Galactocentric distance, while 
for the peculiar clusters it is larger by almost 2 kpc (see the 
concentrations of the corresponding groups in Fig. 3b). The small 
Galactocentric distance at which the maximum density of ''thin-disk'' 
clusters is observed in the diagram suggests that they are formed 
mainly within the solar circle, where the number of massive dense 
interstellar clouds is great (however, they may be simply invisible 
in the Galactic plane at great distances). At the same time, the mean 
ages less than those for the peculiar clusters imply that the Galactic 
clusters generally also live less due to the long-term destructive 
effect from massive clouds of interstellar gas and spiral density 
waves. The scale height for the "thick-disk" clusters was found to be 
approximately a factor of 3.5 larger than that for the Galactic 
clusters and, outside the error limits, larger than that for the old 
($>3$ Gyr) metal-poor ($[Fe/H] < -0.13$) field thin--disk stars, which 
is $220 \pm 20$~pc (see Marsakov and Shevelev 1995). However, it turned 
out to be much smaller than that for the field thick--disk RR Lyr 
stars, which is $0.74 \pm 0.05$~kpc, but, at the same time, the mean 
half--thickness of the ''halo'' clusters ($<Z_{max}> = 7.71$~kpc) turned
out to be even slightly larger than the scale height of the field 
RR Lyr stars of the Galactic accreted halo (see Borkova and Marsakov2002).

{\bf PROPERTIES OF DIFFERENT POPULATIONS OF OPEN CLUSTERS} 

Figure 3a presents the distributions of the ''thick''- and 
''thin--disk'' and ''halo'' clusters as well as field Galactic--disk 
stars from the catalogue by Borkova and Marsakov (2005) on the 
[Fe/H]-[Mg/Fe] plane. As we see, the field stars show a rather narrow 
sequence indicative of their genetic relationship (for more details, 
see Marsakov and Borkova 2006a). The ''thin--disk'' clusters do not go 
outside this sequence and, therefore, they can be assumed to be 
composed mainly of the matter reprocessed in previous generations of 
thin--disk stars, i.e., these stellar objects are genetically related. 
Note that we selected the clusters into this group by their orbital 
elements and metallicities, while here we discuss the magnesium 
abundance. If these clusters were formed from matter that experienced a 
different chemical history, then the [Mg/Fe] ratios in them could go 
well outside the sequence described for the field stars. As we see, the 
''thick--disk'' clusters lie in a slightly wider band than do the field 
stars and are located, on average, slightly above the sequence of field 
thin--disk stars. The field stars of the so-called ''metal--rich wing'' 
($[Fe/H] > -0.4$) of the thick disk behave precisely in this way 
(see, e.g., Bensby et\,al.~2003; Marsakov and Borkova 2005). The ages 
and metallicities for these stars are the same as those for the 
thin--disk stars, while their kinematics is like that for the thick 
disk. Therefore, the debates over their origin have been conducted to 
the present day. The coincidence of both chemical and kinematical 
characteristics for the corresponding populations of clusters and field 
stars suggests that the field stars of the metal-rich wing of the thick 
disk could be the remnants of disrupted ''thick--disk'' open clusters. 
The ''halo'' clusters behave in the diagram quite differently: not only 
that they are, on average, less metal--rich, they exhibit a very large 
spread in relative magnesium abundances, with only one of the five 
clusters being within the band occupied by the field stars. This 
suggests with a high probability that they were born from matter that 
was genetically weakly related to the matter from which most of the 
thin--disk stars were formed. Having investigated the abundances of 
some chemical elements in five distant old open clusters (Berkeley 20, 
Berkeley 21, NGC 2141, Berkeley 29, Berkeley 31), Yong et\,al.~(2005) 
concluded that they resulted from the stimulation of star formation by 
a series of captures of matter from dwarf satellite galaxies occurred 
in the outer Galactic disk at different times. Note that, according to 
our kinematical criterion, four of these clusters fell into the 
''halo'', while one cluster (NGC 2141) with a slightly enhanced 
(compared to the field thin-disk stars) relative magnesium abundance 
fell into the ''thick disk''.

Based on the hypothesis about the dual nature of open clusters, we can 
naturally explain the emergence of the above-described metallicity jump 
when passing to more distant clusters. Figure~3d presents the 
$R_G-[Fe/H]$ diagram, in which the Galactic and peculiar clusters are 
marked by different symbols. Two regression lines were drawn in the 
diagram: one from the Galactic clusters and the other from the peculiar 
ones. Both correlations are insignificant, because $P_N > 10$\,\% for 
both. Comparison with Fig.~2a shows that approximately the same slopes 
of the regression lines for each population turned out to be much 
smaller than those for the entire set of clusters and they are 
separated in metallicity by $\Delta[Fe/H]\approx -0.3$. A similar 
picture is also observed for the vertical metallicity gradient; in the 
$|z|-[Fe/H]$ diagram, both slopes are zero, within the error limits, 
while the regression lines are separated from each other in metallicity 
even slightly more (the corresponding figure is not shown to save space;
see Fig.~2b). Two factors probably contributed to the formation of the 
jump-like pattern of the radial and vertical metallicity gradients: the 
existence of a corotation zone, which caused the chemical evolutions of 
the inner and outer regions of the Galaxy to be independent 
(see Lepine et\,al.~2011), and, most importantly, an active interaction 
of extragalactic fragments (such as high--velocity clouds, globular 
clusters, or dwarf galaxies) with the interstellar medium of the thin disk.
The flattening of the negative radial metallicity gradient with 
increasing cluster age described in the Section ''The Relation between 
the Chemical Composition and Position in the Galaxy'' can also be 
naturally explained in terms of the existence of two groups of open 
cluster populations. Indeed, as we see from Figs.~1a and~1b, the 
relative number of ''kinematically cold'' clusters (to which all our 
Galactic clusters belong) progressively decreases with increasing age 
and, on average, a factor of 2 less metal-rich peculiar clusters 
constitute an increasingly large percentage (see Fig.~3d). Therefore, 
at an old age, the metallicity gradient is recorded only for the 
peculiar group of clusters, whose magnitude is appreciably smaller than 
that obtained from both groups together. When the evolution of the 
metallicity gradient is interpreted, it should be borne in mind that 
the age of any peculiar cluster is related to the individual capture of 
the extragalactic fragment that triggered local star formation by the 
Galaxy and not to the existence of regular star formation caused by the 
motion of spiral density waves. Therefore, the temporal trend of the 
radial metallicity gradient shown by the open clusters is an additive 
function of several factors: the star formation rate at different 
Galactocentric distances, the time of action of extragalactic objects 
on the interstellar medium, the open cluster disruption rates at 
different distances from the Galactic center and plane, and 
observational selection.
A comparative analysis of the relations between the present positions 
and apogalactic orbital radii of clusters and their age can serve as a 
confirmation of the hypothesis that the peculiar clusters result from 
the action of rapidly moving extragalactic fragments on the 
interstellar medium. Indeed, we see from the $R_G-age$ diagram in 
Fig.~4a that there are only three clusters in the catalogue among the 
clusters younger than, say, 15 Myr and farther than $\approx10.5$~kpc 
from the Galactic center. Since optically bright young clusters are 
seen far, this can be only if no clusters at all have been formed 
recently at great distances. Having gained a significant acceleration 
during their birth, the clusters that were formed near the solar circle 
have not yet had time to go far away. Such young clusters must have 
large apogalactic orbital radii, which we see in Fig. 4b, where there 
are more than ten and a half such clusters in the catalogue. Since the 
mean revolution period of the open clusters around the Galactic center 
is $\approx 230$~Myr and reaches 600 Myr or more for some of the 
peculiar clusters (see Wu et al. 2009), the clusters can in no way fill 
the lower right corner of the $R_G$--age diagram in 15 Myr. If the 
clusters were born at the apogalactic radii of their orbits then the 
young clusters in the $R_G$--age panel would be observed near precisely 
their Ra. The highest probability to detect them near the apogalactic 
radii of their orbits will also be at a steady--state random 
distribution of these objects in orbital phases.
A similar situation is also observed when comparing the observed 
positions of the clusters with the maximum distances of their orbital 
points from the Galactic plane. We see from Fig.~4c that among the 
clusters younger than 15 Myr, only one is located higher than 
$\approx 180$~pc above the Galactic plane, while $Z_{max}$ for about 
twenty of these young clusters exceed considerably this height 
(see Fig.~4d). According to the data from the catalogue by 
Wu et\,al.~(2009), the mean time for a peculiar cluster crossing the 
Galactic plane between the opposite points $Z_{max}$ and $Z_{min}$ is 
$\approx46$~Myr (at a maximum period of 125 Myr), i.e., several times 
greater than the constraint of 15 Myr adopted here. The high velocity 
of their passage by massive interstellar clouds and spiral arms helps 
such clusters to survive by being born within the solar circle, where 
the density of the interstellar medium is particularly high near the 
Galactic plane, reducing the duration of their destructive effect. In 
contrast, the peculiar clusters spend much of the time high above the 
Galactic plane and far from the Galactic center, where there are much 
fewer gravitational potential nonuniformities. Thus, a comparative 
analysis of the age dependences of the present positions of open 
clusters and their maximum distances from the Galactic center and plane 
suggests that the bulk of the clusters were formed within a 
Galactocentric radius of $\approx10.5$~kpc and closer than 
$\approx180$~pc from the Galactic plane, and only in the course of time 
did some of them recede to considerable distances.
Thus, the causes of the formation of a particular open cluster are 
highly individual. This led to a wide variety of both external 
spatial--kinematical characteristics and internal chemical and physical 
parameters among them. Moreover, the difference in existence conditions 
led to different lifetimes of the clusters dependent on their Galactic 
orbital elements. As a result, all of the relations between the age, 
chemical composition, and spatial-kinematical characteristics observed 
for them turn out to be noncoincident with those for the field 
thin--disk stars. In other words, it is inappropriate to model the 
chemical and dynamical evolution of the Galactic thin disk by studying 
the integrated properties of open clusters whose stars account for only 
a small fraction of the field stars without allowance for the specific 
conditions of their formation and disruption.
The total number of open clusters in the Galaxy is estimated to be from 
thirty to one hundred thousand (see Portegies Zwart et\,al.~2010; 
Piskunov et\,al.~2006), while we know slightly more than 2000 
(Dias et\,al.~2002, version 3.1); the orbits were found only for 
$\approx 500$ 
of them, while the metallicities were found for fewer than three 
hundred clusters. Therefore, to increase the reliability of the results 
obtained, the estimates, and the conclusions, it is desirable to 
determine the required parameters for the already known clusters and to 
invoke the already published data on the detailed chemical composition 
of stars in the clusters.

\section*{ACKNOWLEDGMENTS}

We are grateful to V.M.~Danilov and A.V.~Loktin for a preliminary 
familiarity with the manuscript and constructive additions. This work 
was supported by the Russian Foundation for Basic Research (project no. 
11-02-00621 a). V.A.~Marsakov also thanks the Ministry of Education and 
Science of the Russian Federation for partial support (project P 685).

\renewcommand{\refname}{Список литературы}

\newpage

\begin{figure*}
\centering
\includegraphics[angle=0,width=0.96\textwidth,clip]{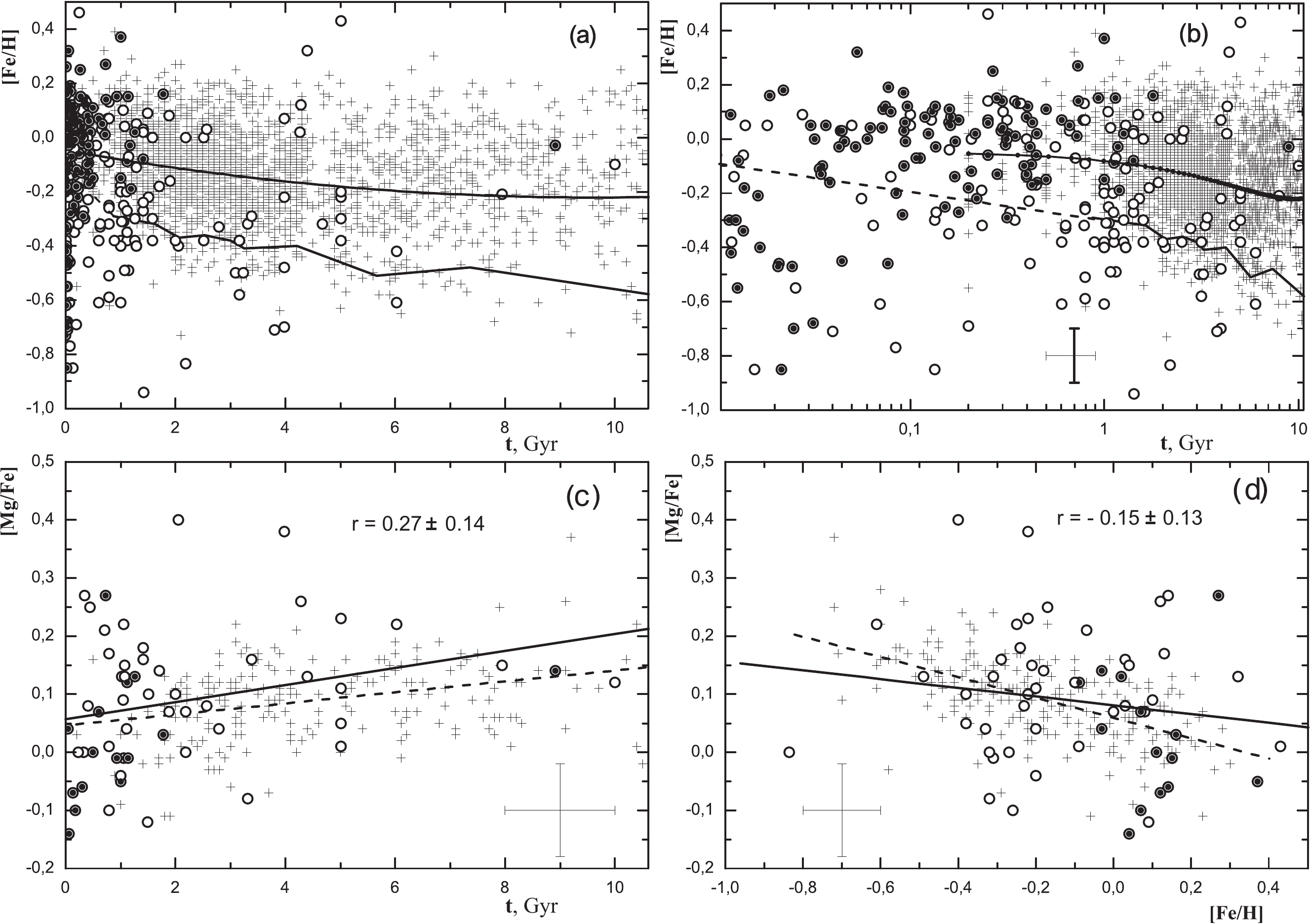}
\caption{(a) Age--metallicity diagram for open clusters and field 
         thin--disk stars; (b) the same but the horizontal axis is on 
         a logarithmic scale; (c) age--[Mg/Fe] diagram; (d) [Fe/H]-[Mg/Fe]
         diagram. The open circles, the dots inside circles and the 
         crosses indicate the open clusters, ''kinematically cold'' 
         clusters, and field thin-disk stars, respectively. In panels 
         (a) and (b), the curved line represents the cubic polynomial 
         fit to the age--metallicity relation for field stars; the 
         polygonal line represents the lower envelope for field stars. 
         In panels (c) and (d), the solid and dashed lines indicate the 
         regression lines for the clusters and field stars, 
         respectively. The mean errors of the parameters are shown.}
\label{fig1}
\end{figure*}

\newpage

\begin{figure*}
\centering
\includegraphics[angle=0,width=0.90\textwidth,clip]{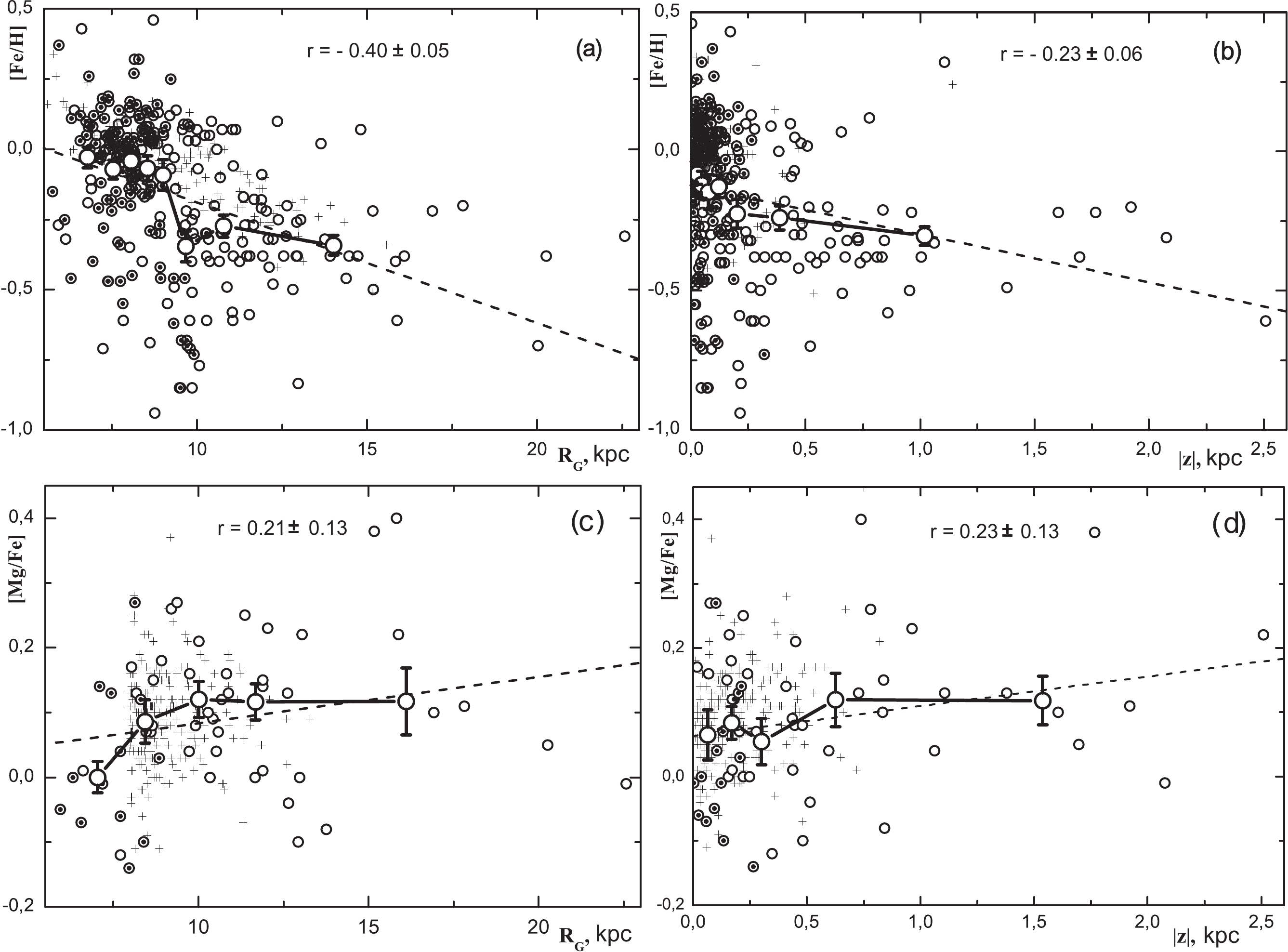}
\caption{Galactocentric distance-metallicity (a), distance from the 
         Galactic plane--metallicity (b), Galactocentric distance--relative
         magnesium abundance (c), and distance from the Galactic 
         plane--relative magnesium abundance (d) diagrams. The 
         designations are the same as those in Fig. 1. In panels 
         (a) and (b), the dashed lines represent the regression lines 
         for the clusters, the open circles with bars connected by the 
         polygonal line are the mean metallicities and their errors in 
         narrow distance ranges.}
\label{fig2}
\end{figure*}

\newpage

\begin{figure*}
\centering
\includegraphics[angle=0,width=0.96\textwidth,clip]{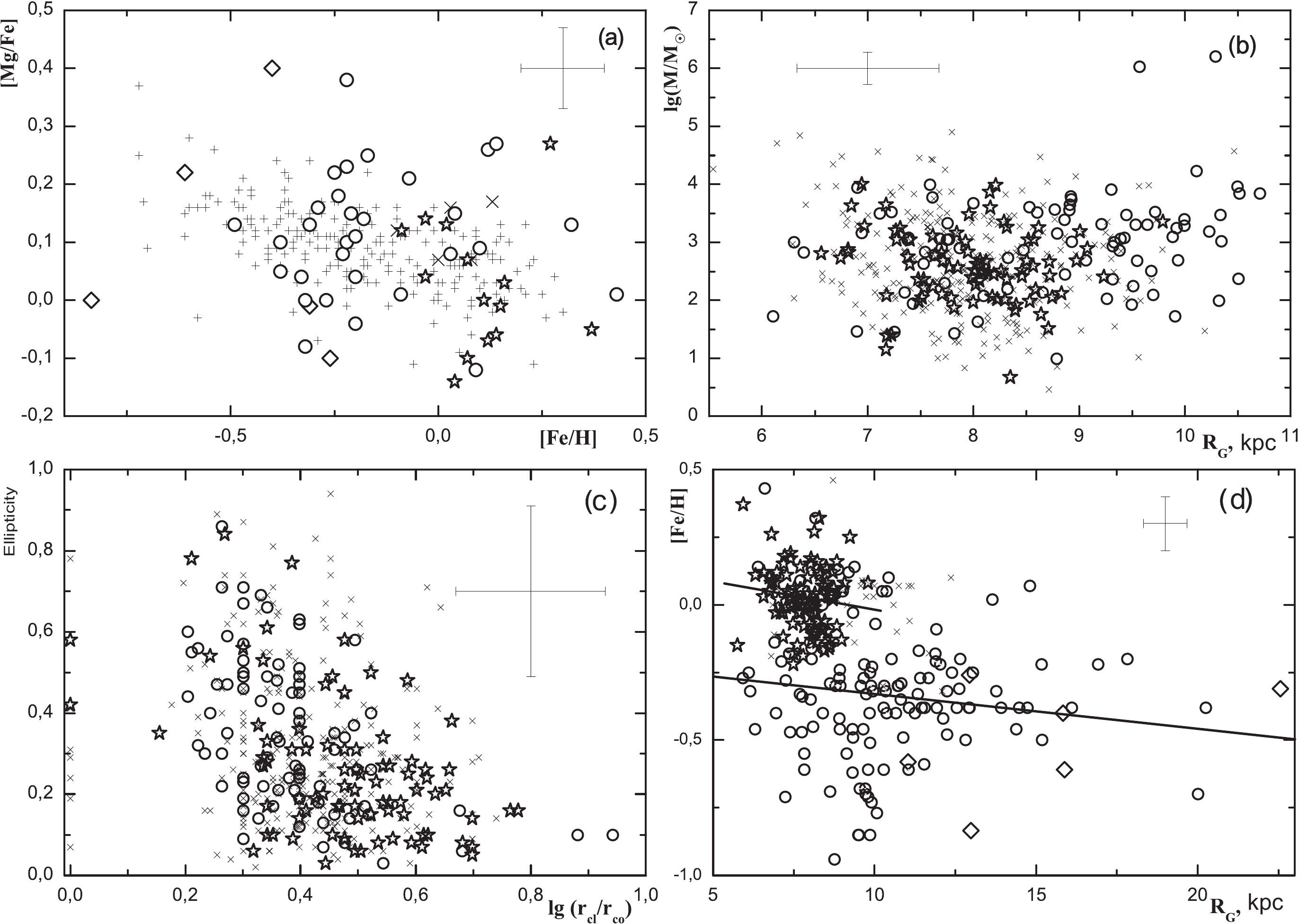}
\caption{Metallicity--relative magnesium abundance (a), Galactocentric 
         distance--mass (b), central concentration-ellipticity (c), and 
         Galactocentric distance-metallicity (d) diagrams for the open 
         clusters. The stars, circles, diamonds, and crosses indicate 
         the ''thin-disk'' clusters, ''thick-disk'' clusters, ''halo'' 
         clusters, and clusters without orbital elements, respectively; 
         the pluses indicate the field thin--disk stars.}
\label{fig3}
\end{figure*}

\newpage

\begin{figure*}
\centering
\includegraphics[angle=0,width=0.96\textwidth,clip]{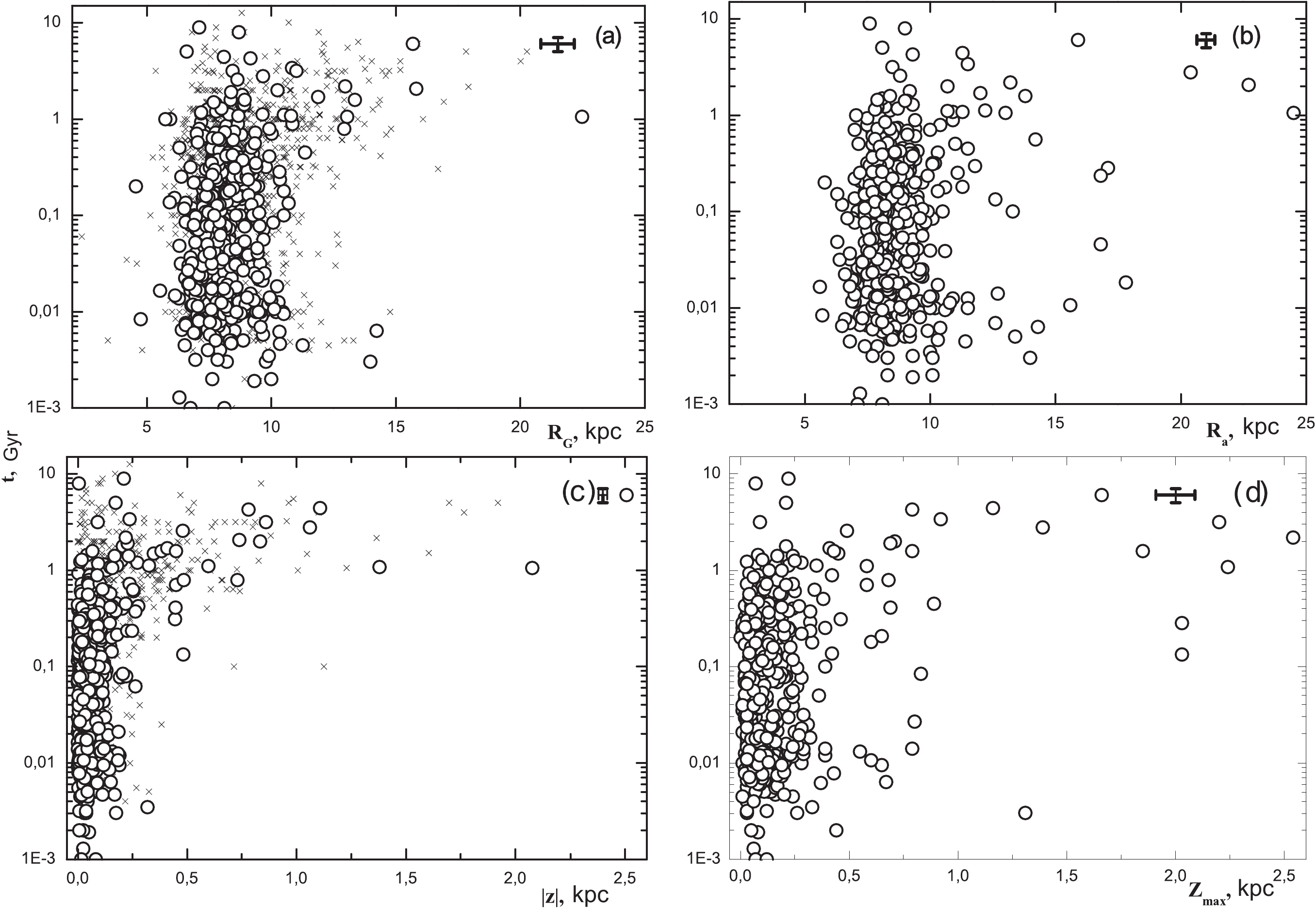}
\caption{Galactocentric distance-age (a), apogalactic orbital 
         radius--age (b), distance from the Galactic plane-age (c), 
         and maximum distance of the orbital points from the Galactic 
         plane--age (d) diagrams for the open clusters. The crosses and 
         the circles indicate all open clusters and those with the 
         orbital elements, respectively.}
\label{fig4}
\end{figure*}


\begin{thebibliography}{}

\bibitem[1.]{1.} G.~Andreuzzi, A.~Bragaglia, M.~Tosi, 
et\,al., MNRAS {\bf 412}, 1265 (2011) 

\bibitem[2.]{2.} S.M.~Andrievsky, D.~Bersier, 
V.V.~Kovtyukh, et al., Astron. Astrophys. {\bf 384}, 140 (2002a)

\bibitem[3.]{3.} S.M.~Andrievsky, R.E.~Luck, 
P.~Martin, et\,al., Astron. Astrophys. {\bf 413}, 159 (2004)

\bibitem[4.]{4.} S.M.~Andrievsky, V.V.~Kovtyukh, 
R.E.~Luck, et\,al., Astron. Astrophys. {\bf 381}, 32 (2002б)

\bibitem[5.]{5.} S.M.~Andrievsky, V.V.~Kovtyukh, 
R.E.~Luck, et\,al., Astron. Astrophys. {\bf 392}, 491 (2002в)

\bibitem[6.]{6.}  W.D.~Arnett, Astrophys. J. {\bf 219}, 1008 (1978).

\bibitem[7.]{7.} T.~Bensby, S.~Feldsing, I.~Lundstrem, 
Astron. Astrophys. {\bf 410}, 527 (2003)

\bibitem[8.]{8.} T.V.~Borkova and V.A.ЁMarsakov, Astron. Rep. {\bf 79}, 
510 (2002)

\bibitem[9.]{9.} T.V.~Borkova and V.A.ЁMarsakov, Astron. Rep. {\bf 49}, 
405 (2005).

\bibitem[10.]{10.} D.~Vande Putte, T.P.~Garnier, 
I.~Ferreras, et\,al., MNRAS {\bf 407}, 2109 (2010)

\bibitem[11.]{11.} Z.-Yu. Wu, X.~Zhou, J. Ma, et\,al., 
MNRAS {\bf 399}, 2146 (2009) 

\bibitem[12.]{12.} M.L.~Gozha, T.V.~Borkova  and V.A.~Marsakov, Astron. 
Lett. {\bf 38}, 519 (2012)

\bibitem[13.]{13.} W.S.~Dias, B.S.~Alessi, A.~Mointinho, et\,al.,
http://www.astro.iag.usp.br/wilton , Astron. Astrophys. {\bf
389}, 871 (2002) 

\bibitem[14.]{14.} D.~Yong, B.W.~Carney, M.L.~Teixera de Almeida,
Astron. J. {\bf 130}, 597 (2005)

\bibitem[15.]{15.} Y.~Yoshii, T.~Tsujimoto, K.~Nomoto, Astrophys.
J. {\bf 462}, 266 (1996)

\bibitem[16.]{16.} V.V.~Koval', V.A.~Marsakov, and T.V.~Borkova, 
Astron. Rep. {\bf 86}, 844 (2009)

\bibitem[17.]{17.} R.E.~Luck, W.P.~Gieren, S.M.~Andrievsky, 
et\,al., Astron. Astrophys. {\bf 401}, 939 (2003)

\bibitem[18.]{18.} L.~Magrini, P.~Sestito, S.~Randich, 
et\,al., Astron. Astrophys. {\bf 494}, 95 (2009) 

\bibitem[19.]{19.} V.A.~Marsakov and T. V. Borkova, Astron. Lett. 
{\bf 32}, 419 (2006а)

\bibitem[20.]{20.} V.A.~Marsakov and T. V. Borkova, Astron. Lett. 
{\bf 32}, 545 (2006б)

\bibitem[21.]{21.} V.A.~Marsakov and T. V. Borkova, Astron. Lett. {\bf 31},
515 (2005).

\bibitem[22.]{22.} V.A.~Marsakov, V. V. Koval', T. V. Borkova, et al., Astron.
Rep. {\bf 55}, 667 (2011).

\bibitem[23.]{23.} V.A.~Marsakov,  Yu.G.~Shevelev,Astron.
Rep. {\bf 72}, 630 (1995)

\bibitem[24.]{24.} V.A.~Marsakov, A.A.~Suchkov, Yu.G.~Shevelev,
 Astrophys. Space Sci. {\bf 172}, 51 (1990)

\bibitem[25.]{25.} F.~Matteucci, I.~Greggio, Astron. Astrophys.
{\bf 154}, 279 (1986)

\bibitem[26.]{26.} A.E.~Piskunov, N.V.~Kharchenko, S.~Roser,
et\,al., Astron. Astrophys. {\bf 445}, 545 (2006)

\bibitem[27.]{27.} S.~Portegies Zwart, S.~McMillan,
M.~Gieles, Ann. Rev., {\bf 48}, 431 (2010)

\bibitem[28.]{28.} F.K.~Thielemann, K.~Nomoto, Y.~Yokio,
Astron. Astrophys. {\bf 158}, 17 (1986)

\bibitem[29.]{29.} B.M.~Tinsley, Astrophys. J. {\bf 229}, 1046 (1979)

\bibitem[30.]{30.} J.~Holmberg, B.~Nordstrem, J.~Andersen,
Astron. Astrophys. {\bf 501}, 94 (2009) 

\bibitem[31.]{31.} L.~Chen, J.L.~Hou, J.J.~Wang, Astron. J. 
{\bf 125}, 1397 (2003)

\bibitem[32.]{32.} L.~Chen, J.L.~Hou, J.L.~Zhao, et \,al., 
A Giant Step: from Milli- to Micro-arcsecond Astrometry, IAU Symp. 248 
(Ed. W.J. Jin, I. Platais, M.A.C. Perryman, Dordrecht: Kluwer Acad. Publ.,
2008), p. 433



\end{thebibliography}
\end{document}